\documentclass[twocolumn,article,11pt]{IEEEtran}
\usepackage{epsf}             %incluir figuras encapsulated postscript
\usepackage{amssymb}
\usepackage{lipsum}
\usepackage[latin1]{inputenc}
\usepackage{amssymb,amstext,amsmath,amsthm}
\usepackage{array}
\usepackage[ruled,vlined]{algorithm2e}
\usepackage{verbatim}
\usepackage{float}
\usepackage[dvips]{graphicx}
\usepackage{tikz}
\usepackage{fancyhdr}
\usetikzlibrary{arrows,automata}
\usepackage{tabularx}
\usepackage{cite}
\usepackage[english]{babel}
\ifCLASSINFOpdf
   \usepackage[pdftex]{graphicx}

   \DeclareGraphicsExtensions{.pdf,.jpeg,.png}
\else

\fi
\usepackage{acronym}
\usepackage{epstopdf}

\begin{document}
% paper title
% can use linebreaks \\ within to get better formatting as desired
%\title{Modeling of Fading Dependent Channels}
\title{Network Coded Handover in IEEE 802.11}
\author{Samah A. M. Ghanem,~\IEEEmembership{Senior Member,~IEEE}\\
%\IEEEauthorblockA{Instituto de Telecomunicaçُes\\
%Faculdade de Engenharia da Universidade do Porto, Portugal\\
%Email: samah.ghanem@fe.up.pt}
%\thanks{M. Shell is with the Department
%of Electrical and Computer Engineering, Georgia Institute of Technology, Atlanta,
%GA, 30332 USA e-mail: (see http://www.michaelshell.org/contact.html).}% <-this % stops a space
%\thanks{J. Doe and J. Doe are with Anonymous University.}% <-this % stops a space
%\thanks{Manuscript received April 19, 2005; revised January 11, 2007.}}
}
\maketitle
\begin{abstract}
\boldmath
We propose a network coded handover of a station moving between two IEEE 802.11 access points (AP). To address such novel proposed framework on a small cell WiFi to WiFi AP handoff, we propose a novel model for the Distributed Coordination Function (DCF) of the WiFi IEEE 802.11 with fixed average contention window. We provide a single packet tranmission model which has been extended to N-packets transmission models with and without fragmentation. We also model the N-packet transmission for the uncoded/coded packets broadcast in order to compare the IEEE 802.11 unreliable to reliable coded broadcast with ACK. We analyze the delay over all, unicast and broadcast transmissions, for the scenario considered with a topology with one WiFi AP before the handover. Capitalizing on the set of models and their corresponding mean completion times (delay), we analyze the performance of different mechanisms. Finally, we provide a novel formulation of the Network Coding on the Edge handover when the station is mobile allowing for the derivation of optimal transmission strategies that can define an optimal time, when to switch to the other AP.
\end{abstract}
\section{	Introduction}
\normalsize
Mobility is an essential key feature which requires special care in network design. Nowadays, there are several approaches attempting to provide optimal mobility management, from different stack perspectives, starting in TCP/IP Layer 2 with wireless and cellular technologies up to TCP/IP Layer 5 with SIP. However, all these approaches introduce some disruption time while the handover is performed. If different technologies are available, several other approaches provide the ability for the mobile terminals to be connected to the different heterogeneous technologies while moving and decreasing the disruption time. Even for the same technology, it is possible to make use of the "make before break" paradigm, where the new connection and configuration in the mobile terminal is prepared before the handover takes place. However, there is still some disconnection while a handover is performed. 

In a different perspective, Network Coding (NC) constitutes a disruptive paradigm that relies on the mixing (coding) of packets at intermediate nodes in the network. NC is  based on the simple concept that intermediate nodes are allowed to re-mix information flows in addition to routing them, being able to improve network capacity. From a receiver's perspective, it is no longer crucial to focus on gathering specific packets, but to gather enough coded packets to recover the original information. From the mobility perspective, this enables the support of recovering from dropped packets during handover through the transmission of some combinations of packets. 

Many analytical models that describe the operational properties in IEEE 802.11 WLANs are derived the last few years. Particularly, such works try to model the primary MAC techniques: the DCF and/or PCF. DCF is a CSMA/CA scheme which implements a binary exponential backoff mechanism. Different models have been devised based on Bianchi's model of the DCF mechanism \cite{Bianchi}, and different proposed schemes to optimize the network capacity or to enhance the QoS have been conducted. In \cite{Cali}, the authors propose an analytical model to study the throughput of a p-persistent IEEE 802.11 protocol. Such protocol selects a backoff window size that balance collision and idle period costs. Other contributions were built on top can be found in \cite{Hu05} and \cite{Oliveira}. Recently, an EDCF that employs a radically different contention window size as compared to DCF of IEEE 802.11 was introduced into the IEEE 802.11e. Therefore, an accurate analytical and theoretical based understanding is crucial to guide the design and improvement of effective schemes. A set of proposals to improve the QoS via introducing minor changes in the mechanisms like fixing the maximum contention window to be equal to the first contention, or limiting the number of retrials for the sake of decreasing the collisions, increasing fairness, or to mitigate the hidden terminal problem are given in,~\cite{ParkHL06} and \cite{ChunShi}. In other works, the authors try to improve the delay performance of the handover scenario focusing on providing mechanisms that minimize the most contributor to the delay, which is the probing delay,~\cite{macHandoff1},\cite{macHandoff2}.

Therefore, we address the handover problem starting with the analytical models associated to the access and transmission mechanisms. In particular, in this paper we propose a new model of the IEEE 802.11 WiFi DCF, we propose a fixed average contention window, and we analyze the delay and throughput under different transmission modes, in particular the unicast and different broadcast transmission modes. We analyze the station mobility under a quasi-static Rayleigh fading scenario. We finally, make a comparison between the broadcast transmission to the unicast one for the uncoded case like in the DCF to the coded case using NC, i.e., coding across the packets.  

It has been shown that for the coded case, the delay is decreased by taking into consideration the degrees of freedom the station has to be able to decode the coded packets. Therefore, we reduce the effect of packet erasures and so reduce the number of retransmissions required in case of failure. Moreover, we reduce delay as well as increase reliability in comparison with uncoded broadcast. This proposed framework adds to the listen then talk DCF a mechanism to stop talking after a certain time \cite{DLucani1}. This serves in scenarios where streaming applications exist and particularly under handover scenarios where decreasing the delay is an endeavor. If cashing is also considered in the new AP, the station who is performing a handover to a new AP will not suffer real time recognizable delays.

In this paper, the proposed network coded handover, or network coding on the edge solution is a key enabler to future fifth generation 5G communication systems with stringent delay requirments. Such solution allows for the reduction on the disruption of the active sessions running during the handover by the ustilization of network coding. In our approach, network coding on the edge will be the key enabler towards ensuring soft handover, providing the correct delivery of all packets in the running session without duplicating the packets in both networks. In our approach, the idea is to use network coding to send additional coded packets, i.e., linear combination of the original packets, towards the mobile terminal performing handover. Firstly, this enhances information transmission before the handover, when the channel quality for both data and feedback is severely degraded. Secondly, it improves the probability of correct reception of the information after the handover, when the exact knowledge of the status of the terminals is unknown. These two steps shall be instrumental in achieving a seamless handover. 

The choice of the number of coded packets to be sent before and after the handover needs to take into account several issues, such as the mobility pattern of the mobile terminal, signal variation in both the previous and the new access networks, handover time and the expected performance degradation while handover is taking place. 

Moreover, the decision of handover with network coding needs also to be addressed, since it is required to evaluate the cost of handover - in terms of signaling and performance degradation - and the additional cost of introducing coding, both in the extra amount of information delivered and its delay in real-time communications. 

In this paper, we focus on the IEEE 802.11 WiFi networks and we consider a handover scenario of one station between two WiFi APs. We particularly focus on the modeling issue, the delay issue, and the evaluation of current technologies with respect to the usage of NC in such scenario, and we provide a novel formulation to characterize the optimal time to switch to the new AP. 

Therefore, we propose new models for the DCF of the IEEE 802.11, we analyze the delay over the unicast and broadcast transmission for a network topology of one AP and one station. We provide a closed-form expression for the expected time to deliver the $N$ packets for the DCF mechanism, with  unicast, the general broadcast, broadcast with ACK for the uncoded and coded transmission. We have shown that coding across packets in an acknowledged broadcast scenario encounters less delays, higher reliability, and higher throughout than for the uncoded broadcast or unicast cases. We propose a new protocol that utilizes network coding to broadcast coded packets to the station performing handover. This new proposed network-coded handover framework will immensely serve if implemented in the current standardized IEEE 802.11 systems. We build upon constraints that take into consideration the distance of the station and the degrees of freedom it owns to decode the received packets before it switches the connection to the next AP. Therefore, we provide a framework under which the QoS over delay sensitive streaming applications can be radically improved.
\section{ The DCF}
\normalsize
The DCF of the 802.11 is based on a contention based mechanism. Each station contends to access the medium and succeeds in its access after a time the medium is sensed to be idle, called the DIFS. This will let the station generate a random backoff window in the range of $[0,..,CW_{min}]$, then perform fragmentation of the MSDU into a set of MPDUs to be transmitted; this indeed serves in increasing the reliability of transmission via per MPDU acknowledgment (ACK). Therefore, after the first contention window, the $MPDU_{1}$ transmitted, if it receives an ACK, the second $MPDU_{2}$ can be directly transmitted after a SIFS, if no ACK is received, the station waits for another DIFS to confirm that the medium is still idle and then generates a random backoff window in the range of $[0,..,2CW_{min}]$, then it generates $MPDU_{1}$ again. Therefore, this backoff mechanism dictates a new contention if a failure in transmission or if no ACK is received until the last level of the backoff $[0,.., CW_{max}]$. If the transmission was successfully established over all the $N$ packets $[MPDU_{1},..,MPDU_{N}]$, the backoff is only done at the beginning and is not repeated along the process, such that the station transmits as follows, DIFS, CW1, $MPDU_{1}$, SIFS, ACK, SIFS, $MPDU_{2}$, SIFS, ACK, SIFS,.., SIFS, ACK, SIFS, $MPDU_{N}$, SIFS, ACK.
\subsection{Modeling a Single-Packet Unicast Transmission in the DCF}
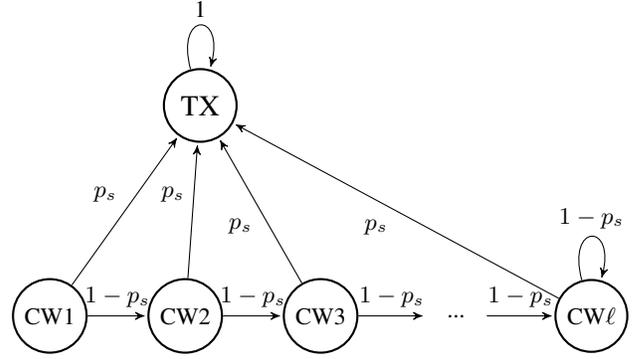
\begin{figure} 
\begin{center}
\begin{tikzpicture}[->,>=stealth',shorten >=1pt,auto,node distance=1.8cm]
\tikzstyle{every state}=[shape=circle,fill=white,draw=black,thick,text=black,scale=1]
\node[state]             (TX) at (3,2) {TX};
%\small
\footnotesize 
\tikzstyle{every state}=[fill=white,draw=black,thick,text=black,scale=1]
\node[state]     (CW1)   [below of=TX]  at (1,1)   {CW1};
\tikzstyle{every state}=[fill=white,draw=black,thick,text=black,scale=1]
\node[state]             (CW2) [right of=CW1] {CW2};
\tikzstyle{every state}=[fill=white,draw=black,thick,text=black,scale=1]
\node[state]             (CW3) [right of=CW2] {CW3};
\tikzstyle{every state}=[fill=white,draw=white,thick,text=black,scale=1]
\node[state]             (CW4) [right of=CW3] {...};
\tikzstyle{every state}=[fill=white,draw=white,thick,text=black,scale=1]
\node[state]             (CW5) [right of=CW3] {...};
\tikzstyle{every state}=[fill=white,draw=black,thick,text=black,scale=1]
\node[state]             (CWl) [right of=CW4] {CW$\ell$};
\path [black] (CW1) edge node {$1-p_s$} (CW2);
\path [black] (CW2) edge node {$1-p_s$} (CW3);
\path [black] (CW4) edge node {$1-p_s$} (CWl);
\path [black] (CW3) edge node {$1-p_s$} (CW5);
\path [black] (CWl) edge [loop above] node {$1-p_s$} (CWl);
\path [black] (CW1) edge node {$p_s$} (TX);
\path [black] (CW2) edge node {$p_s$} (TX);
\path [black] (CW3) edge node {$p_s$} (TX);
\path [black] (CWl) edge node {$p_s$} (TX);
\path [black] (TX) edge [loop above] node {1} (TX);
\node (finish) [left of=TX] {};
\end{tikzpicture}
\end{center}
\normalsize
\vspace{0.1cm}
\caption{Single-packet unicast transmission in the IEEE 802.11 DCF.}
\label{Fig:1}
\end{figure}

\normalsize
Consider the absorption Markov chain shown in Figure~\ref{Fig:1}, this models the DCF mechanism for a single packet unicast transmission. Of particular relevance is to note that reliability in the unicast transmission dictates the usage of a contention mechanism by the station. The station after a DIFS and first backoff window $CW1$ will transmit its first packet with probability of success $p_{s}=(1-p_{e})(1-p_{ack})$, this fact will let the process to be absorbed at the state of transmission TX, where the station can continue to transmit the rest of the packets. However, if the first transmission was not successfully established due to erasure probability $p_{e}$ or to a loss in the ACK with probability $p_{ack}$, the station will choose to move to the next backoff stage with probability $1-p_{s}$, such that it can only access the medium after a new contention with a backoff equal to $CW2$, then it retransmits the first message, if successfully, the Markov chain will be absorbed into the transmission state, if not, the process continues until the last backoff trial at level $\ell$ until its finally absorbed, i.e., the first packet is successfully transmitted. Similarly, the second packet will be transmitted with the same process until the $N$th packet is transmitted from the station. For the sake of simplicity, First: we considered that if the transmission failed with probability $1-p_{s}$ at the last backoff stage, the state stays as is - with a self loop - and is not absorbed to let the station keep trails without being absorbed into a distinct fail state, hence there is no limit defined in the number of retrials, and the maximum contention window is fixed at the last retry to its maximum size. Second: we didn't consider the frozen backoff case, i.e., the case when there is another station that can access the medium before the last decrement of the current station backoff takes place, i.e., if the current station senses a DIFS first, it will have the priority to continue accessing the medium, as long as its all MPDUs - within an MSDU - are to be transmitted; this is similar to the broadcast transmission with almost zero backoff but with higher reliability.
\subsection{The Expected Time to Deliver First Packet}
\normalsize
The model in Figure~\ref{Fig:1} illustrates the delay the packet encounters until it is transmitted at the absorption state TX.  We provide a closed-form expression for the average time to deliver the first packet capitalizing on the time to absorption per transient state. The starting time is the MPDU period $T_{p}$ plus the first contention window $CW1\in[0,...,CW_{min}]$ generated by the station after sensing the medium idle for a DIFS period. The expected time to deliver one packet can be written as the sum over all the expected times to deliver the packet at each contention stage. If the time to deliver the packet at first trial with probability of successful transmission $p_{s}$ is $T_{d}(\omega_{1})$, then this is the time to deliver the first packet. However, if the packet is not transmitted~(absorbed) at first trial, a random contention window $CW2\in[0,...,2CW_{min}]$ is generated, and the expected time to deliver this first packet at second contention stage with probability $p_{s}(1-p_{s})$ is $T_{d}(\omega_{2})$, if the packet is not successfully transmitted at this stage, a random contention window $CW3\in[0,...,3CW_{min}]$ is generated, and the expected time to deliver this first packet at second contention stage with probability $p_{s}{(1-p_{s})}^{2}$ is $T_{d}(\omega_{3})$. As far as the first packet transmission is not successfully established, the backoff mechanism will be repeated such that at the last backoff stage,  the station generates a random contention window $CW\ell\in[0,...,CW_{max}]$, and successful transmission at this stage occurs with an expected time to deliver this first packet with probability $p_{s}{(1-p_{s})}^{\ell-1}$ is $T_{d}(\omega_{\ell})$, the self loop with probability of failure $(1-p_{s})$ at the last contention stage corresponds to the finite number of retrials the station will do until it transmits the packet, so no fail state is considered. Therefore, the average expected time to deliver the first packet starting from the first contention stage is as follows:
\begin{multline}
\mathbb{E}\left[T_{first~packet}\right]=T_{d}(\omega_{1})+(1-p_{s})T_{d}(\omega_{2})+\\{(1-p_{s})}^{2}T_{d}(\omega_{3})+~.~.~.~+\\
{(1-p_{s})}^{\ell-2}T_{d}(\omega_{\ell-1})+\frac{{(1-p_{s})}^{\ell-1}}{p_{s}}T_{d}(\omega_{\ell})
\end{multline}
However, digging into the depths of the equation by breaking down the components of the time to deliver the first packet at each contention stage, will be as follows:
\begin{equation} %	I SUMMARIZED THE SET OF EQUATIONS ABOVE TO THE FOLLOWING ONE SAMAH 12-06-2016
T_{d}(\omega_{i})=\bar{T}_{p}+\mathbb{E}\left[\omega_{i}\right], \forall i=1,...,\ell
\end{equation}
Therefore, we can write the expected time to deliver the first packet in the following closed-form:
\begin{multline}
\label{a12}
\mathbb{E}\left[T_{first~packet}\right]=\sum_{i=0}^{\ell-2} \bar{T}_{p}{(1-p_{s})}^{i}+\\
\sum_{i=0}^{\ell-2} \mathbb{E}\left[\omega_{i+1}\right]T_{slot}{(1-p_{s})}^{i}\\
+\frac{{(1-p_{s})}^{\ell-1}}{p_{s}}(\bar{T}_{p}+ \mathbb{E}\left[\omega_{\ell}\right]T_{slot})
\end{multline}
It is worth to note that the expected backoff time (in $T_{slot}$) is:
\begin{equation}
\mathbb{E}\left[\omega_{\ell}\right]=\frac{1}{\ell+1}\sum_{i=0}^{\ell} i
\end{equation}
And that,
\begin{equation}
\bar{T}_{p}=DIFS+T_{p}+SIFS+ACK
\end{equation}
\subsection{Modeling the Transmission of $N$ Packets in the DCF}
\normalsize
We shall now present a set of transmission schemes and their corresponding models. In particular, we will study different transmission schemes used on top of the DCF at the MAC Layer of the IEEE 802.11. We will study the unicast without fragmentation, unicast with fragmentation, the uncoded broadcast without ACK, and the uncoded broadcast with ACK. Finally, we will dedicate a new section to discuss the network-coded broadcast with ACK.
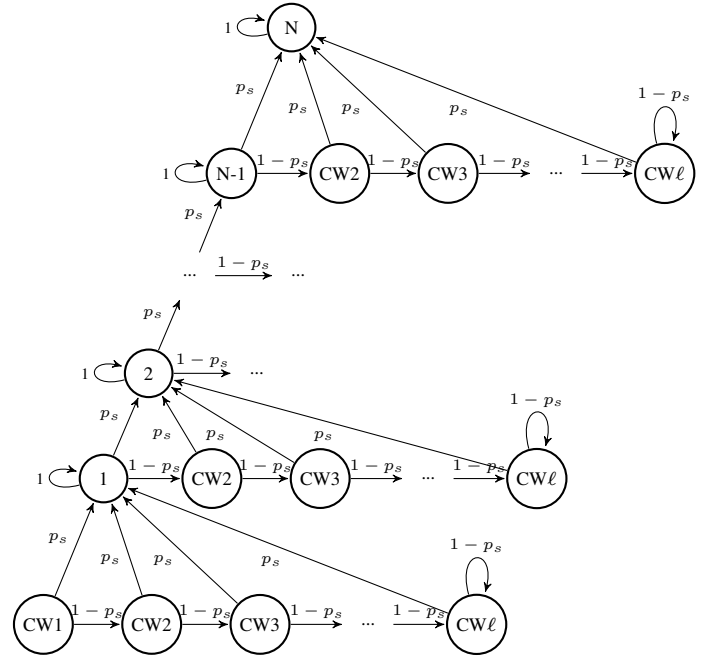
\begin{figure} 
\begin{center}
\tiny
\begin{tikzpicture}[->,>=stealth',shorten >=0.5pt,auto,node distance=1.2cm]
\tikzstyle{every state}=[shape=circle,fill=white,draw=black,thick,text=black,scale=1.2]
\node[state]             (1) at (1.5,1) {1};
\tikzstyle{every state}=[fill=white,draw=black,thick,text=black,scale=1.2]
\node[state]     (CW1)   [below of=1]  at (0.7,0.5)   {CW1};
\tikzstyle{every state}=[fill=white,draw=black,thick,text=black,scale=1.2]
\node[state]             (CW2) [right of=CW1] {CW2};
\tikzstyle{every state}=[fill=white,draw=black,thick,text=black,scale=1.2]
\node[state]             (CW3) [right of=CW2] {CW3};
\tikzstyle{every state}=[fill=white,draw=white,thick,text=black,scale=1.2]
\node[state]             (CW4) [right of=CW3] {...};
\tikzstyle{every state}=[fill=white,draw=white,thick,text=black,scale=1.2]
\node[state]             (CW5) [right of=CW3] {...};
\tikzstyle{every state}=[fill=white,draw=black,thick,text=black,scale=1.2]
\node[state]             (CWl) [right of=CW4] {CW$\ell$};
\path [black] (CW1) edge node {$1-p_s$} (CW2);
\path [black] (CW2) edge node {$1-p_s$} (CW3);
\path [black] (CW4) edge node {$1-p_s$} (CWl);
\path [black] (CW3) edge node {$1-p_s$} (CW5);
\path [black] (CWl) edge [loop above] node {$1-p_s$} (CWl);
\path [black] (CW1) edge node {$p_s$} (1);
\path [black] (CW2) edge node {$p_s$} (1);
\path [black] (CW3) edge node {$p_s$} (1);
\path [black] (CWl) edge node {$p_s$} (1);
\node (finish) [left of=1] {};
%%%%%% second level in blue%%%%%%%%%%%%%%%%%%%%%
\tikzstyle{every state}=[shape=circle,fill=white,draw=black,thick,text=black,scale=1.2]
\node[state]             (2) at (2.1,2.4) {2}; 
%\tikzstyle{every state}=[fill=white,draw=black,thick,text=black,scale=1]
%\node[state]     (TX1)   [below of=TX2]  at (3,2)   {TX1};
\tikzstyle{every state}=[fill=white,draw=black,thick,text=black,scale=1.2]
\node[state]             (N1) [right of=1] {CW2};
\tikzstyle{every state}=[fill=white,draw=black,thick,text=black,scale=1.2]
\node[state]             (N2) [right of=N1] {CW3};
\tikzstyle{every state}=[fill=white,draw=white,thick,text=black,scale=1.2]
\node[state]             (N3) [right of=N2] {...};
\tikzstyle{every state}=[fill=white,draw=white,thick,text=black,scale=1.2]
\node[state]             (N4) [right of=N2] {...};
\tikzstyle{every state}=[fill=white,draw=black,thick,text=black,scale=1.2]
\node[state]             (N5) [right of=N3] {CW$\ell$};
\path [black] (1) edge node {$1-p_s$} (N1);
\path [black] (N1) edge node {$1-p_s$} (N2);
\path [black] (N2) edge node {$1-p_s$} (N3);
%\path [black] (N3) edge node {$1-p_s$} (N4);
\path [black] (N4) edge node {$1-p_s$} (N5);
\path [black] (N5) edge [loop above] node {$1-p_s$} (N5);
\path [black] (1) edge node {$p_s$} (2);
\path [black] (N1) edge node {$p_s$} (2);
\path [black] (N2) edge node {$p_s$} (2);
\path [black] (N5) edge node {$p_s$} (2);
\node (finish) [left of=2] {};
\tikzstyle{every state}=[fill=white,draw=white,thick,text=black,scale=1.2]
\node[state]             (N100) [right of=2] {...};
\path [black] (2) edge node {$1-p_s$} (N100);
\tikzstyle{every state}=[fill=white,draw=white,thick,text=black,scale=1.2]
\node[state]             (N6) at (2.65,3.7) {...};
\tikzstyle{every state}=[fill=white,draw=white,thick,text=black,scale=1.2]
\node[state]             (N200) [right of=N6] {...};
\path [black] (N6) edge node {$1-p_s$} (N200);
\path [black] (2) edge node {$p_s$} (N6);
\tikzstyle{every state}=[shape=circle,fill=white,draw=black,thick,text=black,scale=1.2]
\node[state]             (N7) at (4,7) {N}; 
\tikzstyle{every state}=[fill=white,draw=black,thick,text=black,scale=1.2]
\node[state]     (N8)   [below of=N7]  at (3.2,6.5)   {N-1};
\tikzstyle{every state}=[fill=white,draw=black,thick,text=black,scale=1.2]
\node[state]             (N9) [right of=N8] {CW2};
\tikzstyle{every state}=[fill=white,draw=black,thick,text=black,scale=1.2]
\node[state]             (N10) [right of=N9] {CW3};
\tikzstyle{every state}=[fill=white,draw=white,thick,text=black,scale=1.2]
\node[state]             (N11) [right of=N10] {...};
\tikzstyle{every state}=[fill=white,draw=white,thick,text=black,scale=1.2]
\node[state]             (N12) [right of=N10] {...};
\tikzstyle{every state}=[fill=white,draw=black,thick,text=black,scale=1.2]
\node[state]             (N13) [right of=N12] {CW$\ell$};
\path [black] (N6) edge node {$p_s$} (N8);
\path [black] (N8) edge node {$1-p_s$} (N9);
\path [black] (N9) edge node {$1-p_s$} (N10);
\path [black] (N10) edge node {$1-p_s$} (N11);
\path [black] (N12) edge node {$1-p_s$} (N13);
\path [black] (N13) edge [loop above] node {$1-p_s$} (N13);
\path [black] (N8) edge node {$p_s$} (N7);
\path [black] (N9) edge node {$p_s$} (N7);
\path [black] (N10) edge node {$p_s$} (N7);
\path [black] (N13) edge node {$p_s$} (N7);
%%%%%%% sam
\path [black] (1) edge [loop left] node {1} (1);
\path [black] (2) edge [loop left] node {1} (2);
\path [black] (N7) edge[loop left] node {1} (N7);
\path [black] (N8) edge [loop left] node {1} (N8);
%%%%% sam
\node (finish) [left of=N7] {};
\end{tikzpicture}
\end{center}
\normalsize
\vspace{0.1cm}
\caption{$N$ packet unicast transmission without fragmentation in the IEEE 802.11 DCF.}
\label{Fig:2}
\end{figure}
\subsection{Unicast without Fragmentation}
\normalsize
Further, we adapt the process of the single-packet unicast transmission in an iterative fashion to model the DCF process in transmitting $N$ packets without fragmentation. Figure~\ref{Fig:2} illustrates the unicast transmission of $N$ packets without fragmentation. Worth to note that each stage corresponds to a single-packet transmission, and the last absorption state of each stage corresponds to the initial state to the next-packet transmission until the $N$th packet is delivered, this is due to the fact that after a successful packet transmission, the backoff mechanism is reset and the station has to contend again, so after a DIFS and CW1 it can establish its second packet transmission with the backoff mechanism if any failure is detected, if second packet is successfully transmitted, the station needs again to contend waiting a DIFS and CW1, and so on and so forth until all $N$ packets are delivered. The expected time to deliver the $N$ packets will be an accumulated sum over the average time to deliver each packet. In fact, this calculation is feasible since we are dealing with an average time. Therefore, the expected time to deliver $N$ packets is the number of packets times the expected time to deliver the first packet, as follows:
\begin{equation}
\mathbb{E}\left[T_{N~packets}\right]=N\mathbb{E}\left[T_{first~packet}\right]
\end{equation}
Note that the time to deliver $N$ packets is a non-linear relation in the probabilistic sense. However, it is a linear relation between the times of delivering each individual packet.
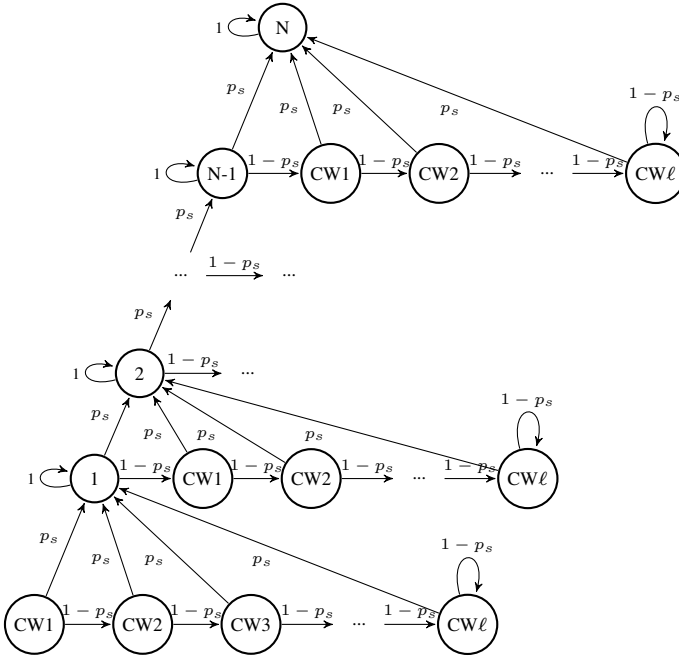
\begin{figure} 
\begin{center}
\tiny
\begin{tikzpicture}[->,>=stealth',shorten >=0.5pt,auto,node distance=1.2cm]
\tikzstyle{every state}=[shape=circle,fill=white,draw=black,thick,text=black,scale=1.2]
\node[state]             (1) at (1.5,1) {1};
\tikzstyle{every state}=[fill=white,draw=black,thick,text=black,scale=1.2]
\node[state]     (CW1)   [below of=1]  at (0.7,0.5)   {CW1};
\tikzstyle{every state}=[fill=white,draw=black,thick,text=black,scale=1.2]
\node[state]             (CW2) [right of=CW1] {CW2};
\tikzstyle{every state}=[fill=white,draw=black,thick,text=black,scale=1.2]
\node[state]             (CW3) [right of=CW2] {CW3};
\tikzstyle{every state}=[fill=white,draw=white,thick,text=black,scale=1.2]
\node[state]             (CW4) [right of=CW3] {...};
\tikzstyle{every state}=[fill=white,draw=white,thick,text=black,scale=1.2]
\node[state]             (CW5) [right of=CW3] {...};
\tikzstyle{every state}=[fill=white,draw=black,thick,text=black,scale=1.2]
\node[state]             (CWl) [right of=CW4] {CW$\ell$};
\path [black] (CW1) edge node {$1-p_s$} (CW2);
\path [black] (CW2) edge node {$1-p_s$} (CW3);
\path [black] (CW4) edge node {$1-p_s$} (CWl);
\path [black] (CW3) edge node {$1-p_s$} (CW5);
\path [black] (CWl) edge [loop above] node {$1-p_s$} (CWl);
\path [black] (CW1) edge node {$p_s$} (1);
\path [black] (CW2) edge node {$p_s$} (1);
\path [black] (CW3) edge node {$p_s$} (1);
\path [black] (CWl) edge node {$p_s$} (1);
\node (finish) [left of=1] {};
%%%%%% second level in blue%%%%%%%%%%%%%%%%%%%%%
\tikzstyle{every state}=[shape=circle,fill=white,draw=black,thick,text=black,scale=1.2]
\node[state]             (2) at (2.1,2.4) {2}; 
%\tikzstyle{every state}=[fill=white,draw=black,thick,text=black,scale=1]
%\node[state]     (TX1)   [below of=TX2]  at (3,2)   {TX1};
\tikzstyle{every state}=[fill=white,draw=black,thick,text=black,scale=1.2]
\node[state]             (N1) [right of=1] {CW1};
\tikzstyle{every state}=[fill=white,draw=black,thick,text=black,scale=1.2]
\node[state]             (N2) [right of=N1] {CW2};
\tikzstyle{every state}=[fill=white,draw=white,thick,text=black,scale=1.2]
\node[state]             (N3) [right of=N2] {...};
\tikzstyle{every state}=[fill=white,draw=white,thick,text=black,scale=1.2]
\node[state]             (N4) [right of=N2] {...};
\tikzstyle{every state}=[fill=white,draw=black,thick,text=black,scale=1.2]
\node[state]             (N5) [right of=N3] {CW$\ell$};
\path [black] (1) edge node {$1-p_s$} (N1);
\path [black] (N1) edge node {$1-p_s$} (N2);
\path [black] (N2) edge node {$1-p_s$} (N3);
%\path [black] (N3) edge node {$1-p_s$} (N4);
\path [black] (N4) edge node {$1-p_s$} (N5);
\path [black] (N5) edge [loop above] node {$1-p_s$} (N5);
\path [black] (1) edge node {$p_s$} (2);
\path [black] (N1) edge node {$p_s$} (2);
\path [black] (N2) edge node {$p_s$} (2);
\path [black] (N5) edge node {$p_s$} (2);
\node (finish) [left of=2] {};
\tikzstyle{every state}=[fill=white,draw=white,thick,text=black,scale=1.2]
\node[state]             (N100) [right of=2] {...};
\path [black] (2) edge node {$1-p_s$} (N100);
\tikzstyle{every state}=[fill=white,draw=white,thick,text=black,scale=1.2]
\node[state]             (N6) at (2.65,3.7) {...};
\tikzstyle{every state}=[fill=white,draw=white,thick,text=black,scale=1.2]
\node[state]             (N200) [right of=N6] {...};
\path [black] (N6) edge node {$1-p_s$} (N200);
\path [black] (2) edge node {$p_s$} (N6);
\tikzstyle{every state}=[shape=circle,fill=white,draw=black,thick,text=black,scale=1.2]
\node[state]             (N7) at (4,7) {N}; 
\tikzstyle{every state}=[fill=white,draw=black,thick,text=black,scale=1.2]
\node[state]     (N8)   [below of=N7]  at (3.2,6.5)   {N-1};
\tikzstyle{every state}=[fill=white,draw=black,thick,text=black,scale=1.2]
\node[state]             (N9) [right of=N8] {CW1};
\tikzstyle{every state}=[fill=white,draw=black,thick,text=black,scale=1.2]
\node[state]             (N10) [right of=N9] {CW2};
\tikzstyle{every state}=[fill=white,draw=white,thick,text=black,scale=1.2]
\node[state]             (N11) [right of=N10] {...};
\tikzstyle{every state}=[fill=white,draw=white,thick,text=black,scale=1.2]
\node[state]             (N12) [right of=N10] {...};
\tikzstyle{every state}=[fill=white,draw=black,thick,text=black,scale=1.2]
\node[state]             (N13) [right of=N12] {CW$\ell$};
\path [black] (N6) edge node {$p_s$} (N8);
\path [black] (N8) edge node {$1-p_s$} (N9);
\path [black] (N9) edge node {$1-p_s$} (N10);
\path [black] (N10) edge node {$1-p_s$} (N11);
\path [black] (N12) edge node {$1-p_s$} (N13);
\path [black] (N13) edge [loop above] node {$1-p_s$} (N13);
\path [black] (N8) edge node {$p_s$} (N7);
\path [black] (N9) edge node {$p_s$} (N7);
\path [black] (N10) edge node {$p_s$} (N7);
\path [black] (N13) edge node {$p_s$} (N7);
%%%%%%% sam
\path [black] (1) edge [loop left] node {1} (1);
\path [black] (2) edge [loop left] node {1} (2);
\path [black] (N7) edge[loop left] node {1} (N7);
\path [black] (N8) edge [loop left] node {1} (N8);
%%%%% sam
\node (finish) [left of=N7] {};
\end{tikzpicture}
\end{center}
\normalsize
\vspace{0.1cm}
\caption{$N$ packet unicast transmission with fragmentation in the IEEE 802.11 DCF.}
\label{Fig:3}
\end{figure}
\subsection{Unicast with Fragmentation} 
\normalsize
When the MSDU size is bigger than a certain limit, the MAC layer do a fragmentation mechanism through which the MSDU is fragmented into a set of MPDUs where the timing between each packet and another is a $SIFS+ACK+SIFS=2SIFS+ACK$. Unlike the Unicast without fragmentation; after a successful MPDU transmission, with $SIFS+ACK$ received, the station continues the $N-1$ MPDUs transmission with $SIFS+ACK+SIFS$ in between. Therefore, the station doesn't have to backoff unless there is a loss in an MPDU or its ACK. This could speed up a long packet transmission with higher reliability. Figure~\ref{Fig:3} illustrates the unicast transmission of $N$ packets with fragmentation. It is worth to notice the shift in timing in the states due to the mechanism discussed. Here the time to deliver the first packet (MPDU) stays the same, while the time to deliver the rest $N-1$ packets will be a little bit different. Therefore, the time to deliver the $N$ packets with fragmentation is the expected time to deliver the first packet plus $N-1$ times the expected time to deliver the second packet, and given by the following closed-form:
\begin{equation}
\mathbb{E}\left[T_{N~packets}\right]=\mathbb{E}\left[T_{first~packet}\right]+(N-1)\mathbb{E}\left[T_{second~packet}\right]
\end{equation}
With,
\begin{multline}
\mathbb{E}\left[T_{second~packet}\right]=\sum_{i=1}^{\ell-1} \bar{T}_{p}{(1-p_{s})}^{i}+\\
\sum_{i=1}^{\ell-1} \mathbb{E}\left[\omega_{i}\right]T_{slot}{(1-p_{s})}^{i}+\\
\frac{{(1-p_{s})}^{\ell}}{p_{s}}(\bar{T}_{p}+ \mathbb{E}\left[\omega_{\ell}\right]T_{slot})+2SIFS+{T}_{p}+ACK
\end{multline}
\subsection{Uncoded Broadcast Transmission without ACK}
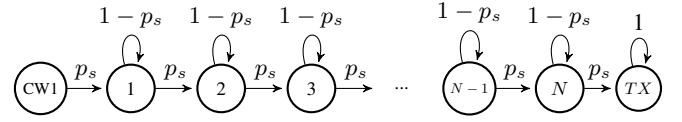
\begin{figure} 
\begin{center}
\begin{tikzpicture}[->,>=stealth',shorten >=1.8pt,auto,node distance=1.5cm]
\footnotesize 
\tikzstyle{every state}=[fill=white,draw=black,thick,text=black,scale=0.7]
\node[state]             (1111) at (1,1) {CW1};
\tikzstyle{every state}=[fill=white,draw=black,thick,text=black,scale=0.8]
\node[state]             (1) [right of=1111] {1};
\tikzstyle{every state}=[fill=white,draw=black,thick,text=black,scale=0.8]
\node[state]             (2) [right of=1] {2};
\tikzstyle{every state}=[fill=white,draw=black,thick,text=black,scale=0.8]
\node[state]             (3) [right of=2] {3};
\tikzstyle{every state}=[fill=white,draw=white,thick,text=black,scale=0.8]
\node[state]             (4) [right of=3] {...};
\tikzstyle{every state}=[fill=white,draw=black,thick,text=black,scale=0.6]
\node[state]             (5) [right of=4] {$N-1$};
\tikzstyle{every state}=[fill=white,draw=black,thick,text=black,scale=0.8]
\node[state]             (6) [right of=5] {$N$};
\tikzstyle{every state}=[fill=white,draw=black,thick,text=black,scale=0.7]
\node[state]             (7) [right of=6] {$TX$};
\path [black] (1111) edge node {$p_s$} (1);
\path [black] (1) edge node {$p_s$} (2);
\path [black] (2) edge node {$p_s$} (3);
\path [black] (3) edge node {$p_s$} (4);
\path [black] (5) edge node {$p_s$} (6);
\path [black] (6) edge node {$p_s$} (7);
\path [black] (1) edge [loop above] node {$1-p_s$} (1);
\path [black] (2) edge [loop above] node {$1-p_s$} (2);
\path [black] (3) edge [loop above] node {$1-p_s$} (3);
\path [black] (5) edge [loop above] node {$1-p_s$} (5);
\path [black] (6) edge [loop above] node {$1-p_s$} (6);
\path [black] (7) edge [loop above] node {1} (7);
\node (finish) [left of=TX] {};
\end{tikzpicture}
\end{center}
\normalsize
\vspace{0.1cm}
\caption{$N$ packets broadcast transmission in the IEEE 802.11}
\label{Fig:4}
\end{figure}

\normalsize
Figure~\ref{Fig:4} illustrates the Markov chain of an $N$ packets uncoded broadcast transmission based on the IEEE 802.11. Broadcast frames neither protected by RTS/CTS, nor acknowledged. Therefore, correct reception cannot be guaranteed. And so, most of the applications use unicast traffic while broadcast is usually used for routing update messages and beacon messages. Unicast favors reliability, while broadcast favors speed.
%\begin{equation}
%%\label{a1}
%\mu_{1}=\mu_{2}+\frac{T_{d}(1)}{p_{s}}\\
%\end{equation}
%
%\vspace{-0.1cm}
%\begin{equation}
%%\label{a2}
%\mu_{2}=\mu_{3}+\frac{T_{d}(2)}{p_{s}}\\
%\end{equation}
%
%\vspace{-0.1cm}
%\begin{equation}
%%\label{a2}
%\mu_{3}=\mu_{4}+\frac{T_{d}(3)}{p_{s}}\\
%\end{equation}
%~~~~~~~~~~~~~~~~~~~~~~~~~~~~~~~~~~~~\vdots
%\begin{equation}
%%\label{a3}
%\mu_{N-1}=\mu_{N}+\frac{T_{d}(N-1)}{p_{s}}\\
%\end{equation}
%
%\vspace{-0.1cm}
%\begin{equation}
%%\label{a4}
%\mu_{N}=\frac{T_{d}(N)}{p_{s}}\\
%\end{equation}

In a similar analytical way to the previous one for the unicast case, we can derive the expected time to deliver $N$ packets via uncoded broadcast transmission for the model in Figure~\ref{Fig:4}. Therefore, the expected time to deliver $N$ packets via broadcast transmission is as follows:
\vspace{-0.2cm}
\begin{equation}
%\label{a4}
\mathbb{E}\left[T_{N~packets}\right]=\sum_{i=1}^{N} \frac{T_{d}(i)}{p_{s}}\\
\end{equation}
\vspace{-0.2cm}
However,
\begin{equation}
%\label{a4}
%{p_{s}}=\binom {i}{j}(1-{p_{e}})^{i}p_{e}^{i-j}\\
{p_{s}}=(1-{p_{e}})\\
\end{equation}
Due to the fact that no backoff mechanism in the broadcast transmission, then, $T_{d}(1)=DIFS+CW1+{T}_{p}$, and the spacing in time between each packet and the other is the $SIFS$ period without any $ACK$, we can rewrite the expected time to deliver $N$ packets for the uncoded broadcast without ACK in a more compact form as follows:
%\vspace{-0.6cm}
%\renewcommand{\theequation}{9.15}
\begin{equation}
%\label{a4}
%\mathbb{E}\left[T_{N-packets}]=\frac{N{T}_{p}+(N-1)SIFS+DIFS+CW1}{(1-{p_{e}}^{N})}\\
\mathbb{E}\left[T_{N~packets}\right]=\frac{N{T}_{p}+(N-1)SIFS+DIFS+CW1}{(1-{p_{e}})}\\
\end{equation}
%%%%%%%%%%%%%%%%%%%%%%%%%%%%%%%%%%%%%%%%%%%%%%%%%%%%%%%%%%%%%%%%%%
In a best case scenario where no erasures happen and all transmission is established after first contention, the expected time to deliver $N$ packets by a station that have already win the medium via broadcast, unicast without fragmentation, and unicast with fragmentation respectively are as follows:
\vspace{-0.2cm}
\begin{equation}
%\label{a4}
\mathbb{E}\left[T_{N-BC}\right]=N{T}_{p}+(N-1)SIFS+DIFS+CW1
\end{equation}

\vspace{-0.72cm}
\begin{equation}
%\label{a4}
\mathbb{E}\left[T_{N-UNIC}\right]=N{T}_{p}+NSIFS+NACK+DIFS+CW1
\end{equation}

\vspace{-0.72cm}
\begin{multline}
\mathbb{E}\left[T_{N-FRAG-UNIC}\right]=N{T}_{p}+\\
(2N-1)SIFS+NACK+DIFS+CW1
\end{multline}
%%%%%%%%%%%%%%%%%%%%%%%%%%%%%%%%%%%%%%%%%%%%%%%%%%%%%%%%%%%%%%%%%%%%
It is quite clear the difference between the delay using the different modes of transmission in the IEEE 802.11, in a best case scenario the difference is at least $NACK+SIFS$ for the non-fragmented case. Notice that we compare between the fragmented and the non-fragmented unicast in a common bases; however, it is worth to note that ${T}_{p}$ of an $MPDU$ is a fraction of a ${T}_{p}$ of a non-fragmented packet, but, we used the same notation for ease of exploitation. In addition, in a real world scenario with a backoff mechanism with random (non-fixed) contention window, the unicast adds huge differences into the delay. Therefore, it is worth to propose a hybrid approach where reliability and speed can be taken into consideration. Therefore, we can consider one example, like using acknowledged broadcast as will be shown in the next section. 
%\Huge
%\setcounter{secnumdepth}{0}
\subsection{Uncoded Broadcast with ACK}
Figure~\ref{Fig:4} partially models a broadcast transmission with an ACK at the end of the transmission, such that two states model the framework. One state represents the batch of packets that need to be broadcasted and acknowledged and the other is the absorption state. However, the probability to successfully deliver $N$ packets differs in the ACK component; where acknowledging a batch of $N$ packets will let the probability of success $p_s=(1-p_e^N)(1-p_{ack})$.  Therefore, the expected time to deliver $N$ packets using broadcast transmission with $ACK$ and without considering the time required for retransmission is as follows:
\begin{equation}
%\label{a4}
\mathbb{E}\left[T_{N~packets}\right]=\frac{N{T}_{p}+NSIFS+DIFS+CW1+T_{w}}{(1-p_{e}^{N})(1-p_{ack})}\\
\end{equation}
%\normalsize
Where $T_{w}=ACK+T_{rt}$ corresponds to the time for acknowledgment and round trip time (RTT).
\section{ Network-Coded Broadcast}
\normalsize
Mixing the two approaches of reliable transmission and fast transmission available in uncoded unicast and uncoded broadcast, we can develop a network-coded broadcast model to have coding across packets and an ACK at the end of the $N$ coded packets; such that if an erasure was detected on one or more of the packets we can retransmit the packet(s). In a similar but more efficient approach to a packet repetition framework, we propose a network-coded approach where coding across packets will be considered. We can transmit a number of coded packets per transmission with linear combinations between the packets over some defined Galois Field (GF). This way the number of transmissions required to restore more packets at the receiver side is less. For example, if we transmit coded packet $aMPDU_{1}$ $\oplus$ $bMPDU_{2}$ $\oplus$ $cMPDU_{3}$ between state 1 and state 2, then we transmit $dMPDU_{1}$ $\oplus$ $eMPDU_{2}$ $\oplus$ $fMPDU_{3}$ between state 2 and state 3, this means that we can save one transmission in the uncoded broadcast case such that we can transmit three packets in two transmissions, such that the receiver needs only to decode the three packets solving the two linear equations by a Gaussian elimination process. %Similar, the probability of successful decoding will be equally-likely for all the packets. 
In turn, for a network-coded broadcast scheme, the probability of successfully transmitting $N_{c}$ coded (or uncoded) packets is as follows:
%%\renewcommand{\theequation}{9.20}
%\begin{equation}
%{p_{s_{i\to j}}}=\sum_{j=1}^{N_{c}}\binom {N_{c}}{j}(1-{p_{e}})^{j}p_{e}^{N_{c}-j}(1-p_{ack})\\
%\end{equation}
%\renewcommand{\theequation}{9.20}
\begin{equation}
{p_{s}}=\sum_{j=1}^{N_{c}}\binom {N_{c}}{j}(1-{p_{e}})^{j}p_{e}^{N_{c}-j}(1-p_{ack})\\
\end{equation}
Its worth to notice that the expected time to deliver $N_{c}$ coded packets is exactly similar to the one for broadcast with an ACK at the end of $N$ transmitted packets. If we are transmitting $N_{i}$ coded packets the probability of successful transmission will be $(1-{p_{e}}^{N_{i}})(1-p_{ack})$. However, there are practical considerations that the previous model lacks: On the one hand, the previous model doesn't consider the time required to retransmit the uncoded or coded packets that are not successfully acknowledged. On the other hand, from a network coding perspective, the previous model doesn't take into consideration the degrees of freedom the receiver owns to be able to decode successfully the received packets. Additionally, for a repeated packets framework, the redundancy and the resources wasted are huge since there is a possibility to receive and decode successfully all packets from the first few trials.
Therefore, we develop the network-coded broadcast model similar to the one introduced in \cite{DLucani1} which can be integrated to the DCF functionality of the MAC layer of the IEEE 802.11. 

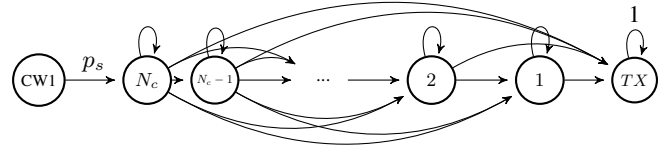
\begin{figure} 
\begin{center}
\begin{tikzpicture}[->,>=stealth',shorten >=2.8pt,auto,node distance=1.8cm]
\footnotesize 
\tikzstyle{every state}=[fill=white,draw=black,thick,text=black,scale=0.7]
\node[state]             (1111) at (0,0) {CW1};
\tikzstyle{every state}=[fill=white,draw=black,thick,text=black,scale=0.8]
\node[state]             (1) [right of=1111] {$N_{c}$};
\tikzstyle{every state}=[fill=white,draw=black,thick,text=black,scale=0.5]
\node[state]             (2) [right of=1] {$N_{c}-1$};
\tikzstyle{every state}=[fill=white,draw=white,thick,text=black,scale=0.8]
\node[state]             (3) [right of=2] {...};
\tikzstyle{every state}=[fill=white,draw=black,thick,text=black,scale=0.8]
\node[state]             (4) [right of=3] {$2$};
\tikzstyle{every state}=[fill=white,draw=black,thick,text=black,scale=0.8]
\node[state]             (5) [right of=4] {$1$};
\tikzstyle{every state}=[fill=white,draw=black,thick,text=black,scale=0.7]
\node[state]             (6) [right of=5] {$TX$};
\path [black] (1111) edge node {$p_s$} (1);
\path [black] (1) edge node {} (2);
\path [black] (2) edge node {} (3);
\path [black] (3) edge node {} (4);
\path [black] (4) edge node {} (5);
\path [black] (5) edge node {} (6);
\path [black](1)  edge [bend right] node {}  (4);
\path [black](2) edge [bend right] node {} (4);
\path [black](1) edge [bend right] node {} (5);
\path [black](2) edge [bend right] node {} (5);

\path [black](1) edge [bend left] node {} (3);
\path [black](2) edge [bend left] node {} (3);
\path [black](1) edge [bend left] node {} (6);
\path [black](2) edge [bend left] node {} (6);
\path [black](4) edge [bend left] node {} (6);

\path [black] (1) edge [loop above] node {} (1);
\path [black] (2) edge [loop above] node {} (2);
\path [black] (4) edge [loop above] node {} (4);
\path [black] (5) edge [loop above] node {} (5);
\path [black] (6) edge [loop above] node {1} (6);
\node (finish) [left of=TX] {};
\end{tikzpicture}
\end{center}
\normalsize
\vspace{0.1cm}
\caption{$N$ coded packets broadcast transmission.}
\label{Fig:5}
\end{figure}

Figure~\ref{Fig:5} illustrates the network-coded broadcast model through which the transmission process is adaptive to the receiver experience. First, we transmit a linear combination of $N_{c}$ coded packets, where those packets correspond to a linear combination of the $MPDU_{1}$ $\oplus$...$\oplus$ $MPDU_{c}$. If the $N_{i}$ coded packets are received successfully with probability $(1-{p_{e}}^{N_{i}})(1-p_{ack})$, the chain is absorbed. If a failure in transmission occurs at state $i$, a self transition will occur with probability $1-(1-{p_{e}}^{N_{i}})(1-p_{ack})$. If not, a transition will occur with probability $p_{i\to j}$. Thus, if a packet erasure occurs and the receiver received less number of packets; the receiver will send back to the transmitter - via the ACK message - an information about the degrees of freedom it requires, such that the chain transition between state $i$ and $j$ corresponds to the number of received packets $i-j$, and the state to where the transition happens corresponds to the coded packets need to be re-transmitted to the receiver, therefore, the transition probability is not fixed over the Markov chain and the probability of success corresponds to the successful transmission of the ${N_{i}}$ coded packets at the current state $i$. 

The probability of transition from state $i$ to state $j$ for the Markov chain in Figure~\ref{Fig:5} is given by:
%tion from state $i$ to state $j$ for the Markov chain in Figure~\ref{Fig:5} is given by,
%\renewcommand{\theequation}{9.21}
\begin{equation}
{p_{i\to j}}=\binom {N_{i}}{i-j}(1-{p_{e}})^{i-j}p_{e}^{N_{i}-i+j}(1-p_{ack})\\
\end{equation}
And the probability of failure in transmission of ${N_{i}}$ coded packets at state $i$ represented by a self loop is given by:
\begin{equation}
{p_{i\to i}}=(1-{p_{ack}})p_{e}^{N_{i}}+p_{ack}\\
\end{equation}
%We will derive the expected time to deliver $N_{c}$-coded-packets via broadcast transmission for the model in Figure\ref{Fig:5} as follows,
%\begin{equation}
%%\label{a1}
%\mu_{1}=\frac{T_{d}(1)}{p_{1\to 1}}\\
%\end{equation}
%
%\vspace{-0.1cm}
%\begin{equation}
%%\label{a2}
%\mu_{2}=\frac{T_{d}(2)}{p_{2\to 2}}+\frac{p_{2\to 1}\mu_{1}}{p_{2\to 2}}\\
%\end{equation}
%
%\vspace{-0.1cm}
%\begin{equation}
%%\label{a2}
%\mu_{3}=\frac{T_{d}(3)}{p_{3\to 3}}+\frac{p_{3\to 2}\mu_{2}}{p_{3\to 3}}\\
%\end{equation}
%~~~~~~~~~~~~~~~~~~~~~~~~~~~~~~~~~~~~\vdots
%\begin{equation}
%%\label{a3}
%\mu_{N_{c}-1}=\frac{T_{d}(N_{c}-1)}{p_{N_{c}-1\to N_{c}-1}}+\frac{p_{N_{c}-1\to N_{c}-2}\mu_{N_{c}-2}}{p_{N_{c}-1\to N_{c}-1}}\\
%\end{equation}
%
%\vspace{-0.1cm}
%\begin{equation}
%%\label{a4}
%\mu_{N_{c}}=\frac{T_{d}(N_{c})}{p_{N_{c}\to N_{c}}}+\frac{p_{N_{c}\to N_{c}-1}\mu_{N_{c}-1}}{p_{N_{c}\to N_{c}}}\\
%\end{equation}

Therefore, the expected time to deliver the $N_{c}$ coded packets for the model in Figure~\ref{Fig:5} can be written as follows:
%\small
%\renewcommand{\theequation}{9.23}
\begin{equation}
%\label{a4}
\mathbb{E}\left[T_{N_{c}-packets}\right]=T_{i}+\frac{\sum_{j=1}^{i-1}\binom{N_{i}}{i-j}(1-{p_{e}})^{i-j}p_{e}^{N_{i}+i-j}T_{j}}{(1-{p_{e}}^{N_{i}})(1-{p_{ack}})}\\
\end{equation}
%\normalsize

With,
\begin{equation}
T_{i}=\frac{N_{i}{T}_{p}+CW{1}+DIFS+T_{w}+ N_{i}SIFS}{(1-{p_{e}}^{N_{i}})(1-{p_{ack}})},\\
\end{equation}
and,
\begin{equation}
T_{j}=\frac{N_{j}{T}_{p}+CW{1}+DIFS+T_{w}+N_{j}SIFS}{(1-{p_{e}}^{N_{j}})(1-{p_{ack}})}\\
\end{equation}

%Note that $T_{w}=ACK+T_{rt}$ corresponds to the time for acknowlegment and round trip time,
Note that if the transmission is established successfully from the first transmission then $N_{i}=N_{c}$, if after the second transmission then $N_{j}+N_{i}=N_{c}$, and so on and so forth until the receiver receives and decodes successfully all coded packets\footnote{Note that the condition $N_{j}+N_{i}=N_{c}$ is sufficient for the transmission of the $N_{c}$ but not necessary since the station can adaptively account for the erased packets with another packets that matches the degrees of freedom of the receiver.}. Therefore, if the optimal number of packets were transmitted, we can guarantee the minimum time to deliver the packets; and therefore the maximum throughput received, as we will introduce in the following subsection. Notice also that in the framework with network coding for time varying channels, or with adaptive network-coded transmission schemes as those proposed in \cite{samahNC}, \cite{samahNC2},\cite{samahNC3}, we can optimize the number of coded packets per transmission according to the awareness of the estimated channel coefficients and their corresponding erasures. However, in this setup we consider a network with fixed erasure probability.

%\Huge
%\setcounter{secnumdepth}{0}
\subsection{Maximizing Throughput with Optimal Number of Coded Packets}
\normalsize
Our objective is to maximize the throughput via minimizing the delay over all the number of coded packets to be transmitted, which is equivalent to minimizing the expected time to deliver the packets we have:%$1/\mathbb{E}\left[T_{N_{c}-packets}\right]$ 
\begin{equation}
\min_{N_{1},...,N_{c}}{\mathbb{E}\left[T_{N_{i}~packets}\right]}
\end{equation}
This can be solved by optimizing jointly over all coded packets, in that case the optimal number of packets in each transmission can be any number that maximizes the throughput. Otherwise, this can be solved iteratively such that the optimal $N_{1}$ that solves $\min_{N_{1}}T_{1}$ is used to compute $\min_{N_{1},N_{2}}T_{2}$, and so on and so forth. However, if we constrained the number of coded packets need to be transmitted, lets say $N_{1}+N_{2}+...+N_{c}=N_{K}$, this will lead to optimal set with optimized $K$, i.e., with a limited number of re-transmissions to deliver $N_{c}$ coded packets. Notice that a closed-form solution of the optimal number of packets is not possible to find, so the iterative solution is the only way, were a closed-form expression is only possible for ${N_{1}}^{\star}$. 

\section{Slow Fading Handover}
One source of degradation of the signal power over the transmission path is due to fading. Fading occurs due to the path loss of the signal as a function of the distance and shadowing. When the mobile moves through a large distance within the coverage of an AP or at the point of handover, this is characterized by a slow fading process with an average path loss between the AP and the station in fixed locations. However, if the station moves through small distances within the coverage of an AP, this is characterized by fast fading process where rapid variations would occur to the signal levels due to interference of multipaths. In fact, we focus here in the slow fading process to our interest in the scenario where the station is approaching the cell edge and just at the time before initiating the process of handover to another AP. In particular, if the time to deliver the $N$ packets to the mobile terminal at such point is related to the SNR and so to the distance, we can characterize an optimal time when to switch to the new AP, and also when the new AP would start to send the mobile the required coded packets to keep its streaming with the least delay effects. In the simple framework of an uncoded modulated signal (i.e., no forward error correction is considered), a detection of one bit error in the received packet corresponds to one packet loss. Therefore, we capitalize on the relation between the probability of erasure of a packet $p_{e}$ and the bit error probability $p_{b}$, which is defined as:
\begin{equation}
p_{e}=1-(1-{p_{b}})^{B}
\end{equation}
Where B denotes the length (in bits) of a single packet. 

Moreover, the bit error probability $p_{b}$ is well defined in the literature with respect to the SNR or mainly to the energy per bit to the noise ratio for coherent as well as the non-coherent detection. For the uncoded BPSK modulation with coherent detection and without fading, the bit error probability $p_{b}$ is given by:
\begin{equation}
p_{b}=Q(\sqrt[~]{2snr})
\end{equation}
For the uncoded BPSK with coherent detection under flat (one path) Rayleigh fading with random channel gain $h\sim\mathcal{CN}(0,1)$, see \cite{VeliMammela}, is given by:
\begin{equation}
p_{b}=\mathbb{E}\left[Q\left(\sqrt[~]{2\left|h\right|^{2}snr}\right)\right]=\frac{1}{2}\left(1-\sqrt[~]{\frac{snr}{1+snr}}\right)
\end{equation}
However, for the coded BPSK modulation with coherent detection, the bit error probability $p_{b}$ depends on the coding, and the error correction capability of the code, as well as the type of decoding with soft or hard decision. Several upper bounds have been presented for the bit error probability for convolutional codes in~\cite{EsaMalkamaki},\cite{FrengerO}, \cite{PerezCostello}, and \cite{PJLee}. However, we used the upper bound of \cite{EsaMalkamaki} since it is more consistent than other bounds. In the case of IEEE 802.11, where 1/2 rate convolutional codes are used, the pairwise error probability $p_{2}(\delta)$ for coherent detection without fading is given by:
\begin{equation}
p_{2}(\delta)=Q(\sqrt[~]{2\delta r_{c}snr})
\end{equation}
For the convolutional coded BPSK with coherent detection and under flat (one path) Rayleigh fading with random channel gain, see \cite{EsaMalkamaki}, is given by:
\begin{multline}
p_{2}(\delta)=\int_{h}p_{2}(\delta|h)p(h)dh\\=\mathbb{E}\left[Q\left(\sqrt[~]{2\left|h\right|^{2}\delta snr}\right)\right]=\frac{1}{2}\left(1-\sqrt[~]{\frac{\delta snr}{1+\delta snr}}\right)
\end{multline}
Where $\delta$ is the hamming distance of the convolutional code and $r_{c}=k_{c}/n_{c}$ is the code rate.
Therefore, the bit error probability is upper bounded as in \cite{EsaMalkamaki}:
\begin{equation}
p_{b}\leq \sum_{\delta=d_{free}}^{\infty}\frac{c(\delta)}{k_{c}}\int_{h}p_{2}(\delta|h)p(h)dh\\=\sum_{\delta=d_{free}}^{\infty}\frac{c(\delta)}{k_{c}}p_{2}(\delta)\\
\end{equation}
With,
\begin{equation}
c(\delta)=\sum_{w_{i}=1}^{\infty}w_{i}a_{i}(\delta),
\end{equation}
and,
\begin{equation}
a(\delta)=\sum_{w_{i}=1}^{\infty}a_{i}(\delta)
\end{equation}
And $p(h)$ is the probability density function of the Rayleigh fading channel distribution. The distance spectrum of a convolutional code is defined by $c(\delta)$ which corresponds to the error weight in information bits, and $a_{i}(\delta)$ is the number of error events with length $\delta$, and $w_{i}$ bit errors. Note that the free distance $d_{free}$\footnote{The free distance $d_{free}$ is the minimal hamming distance between different encoded sequences, through which the correcting capability of a convolutional code $t$ is upper bounded by $\frac{d_{free}-1}{2}$.} provides a first order asymptotic approximation of the error performance. Therefore, as much as we sum over higher order terms, the upper bound would approach the error performance of the uncoded case, and as the constraint length is higher for good convolutional codes, the number of possible codewords grows exponentially, therefore, its somehow sufficient to sum over terms up to $d_{free}+K$, where $K$ is the constraint length, and this would be a condition of truncation of a path in the decoding process, see \cite{FrengerO}, and \cite{APMPoel}. In the case of IEEE 802.11, where $r_{c}=1/2$ rate convolutional code with generator matrix [133, 171] in octets, the minimum hamming weight of the codewords is $\delta=d_{free}=10$, and the constraint length  $K=7$ corresponding to $2^{K-1}=64$ states, see \cite{IEEE80211}, the distance spectrum is as follows, $a(\delta)=[11,0,38,0,193,0,1331,0,...]~ $and~ $c(\delta)=[36,0,211,0,1404,0,11633,0,...]$, notice that the vector starts at $\delta=d_{free}$, see Table 1 in \cite{FrengerO}. Notice also that the meaning of such setup leads to a formulation of a transfer function of the convolutional code; that presents that at $\delta=d_{free}$, 11 error events of weight 36 occurs and at $\delta=d_{free}+1$, 0 error events of weight 0 occurs, and so on and so forth. This can be easily found by Matlab built-in function (distspec).

\subsection{Free Path Loss}
\normalsize
The average received signal to noise power ratio $snr=P_r/N_{0}$ per symbol, and $Q(.)$ is the complementary cumulative distribution function of a Gaussian random variable. Meanwhile, we will use the free-space propagation model to relate the distance of the mobile to the AP. Let $d$ denote the distance in meters between the AP and the station. $\eta$ is the path loss exponent and $\eta=2$ for free space; i.e., it is environment dependent. In fact, free space path loss is a simple model of propagation. Moreover, it is a convenient choice to consider the channel between the AP and the station as quasi-static Rayleigh fading channel in a handover scenario where the station is approaching the edge of the AP coverage. This type of channel exhibits slow fading and so the fading coefficients remain constant during the transmission of the entire $N$ packets, while changes randomly and independently between different transmission/retransmission according to a complex Gaussian distribution with variance equals to the $snr=c{d}^{-2}$, therefore, $h\sim\mathcal{CN}(0,c{d}^{-2})$. $c$~corresponds to a constant that can be chosen to maintain a given $snr(dB)$ at a given distance. 

The expected time to deliver the $N$ packets for the unicast, broadcast without ACK, and broadcast with ACK transmission modes under the previous assumptions can be directly derived with respect to the SNR and/or the distance by substituting into the probability of erasure. In addition, it~is straightforward to relate the time to deliver the $N$ packets and the throughput to the mobile station velocity via the basic distance-velocity-time relation; $d=tv$, so if we know that the mobile station is moving with velocity $v$ meters/sec, we will know that at time $t$, it will be at location $d$, which means that the expected time required to deliver $N$ packets for the station at this location is $T_{AP1}$ and so, we can search for the optimal number of packets to transmit before switching to the other AP, as well as the prospective new AP can compare its $T_{AP2}$. Therefore, both APs can minimize the time the station requires for the scanning in the handover process by establishing the connection and transmitting cooperatively the coded packets required (cashed) and/or the native (uncoded) ones. 
%\Huge
%\setcounter{secnumdepth}{0}
\section{ Network Coding on the edge: Network Coded Handover}
\normalsize
In this section we will propose an optimal handover scheme based on reliable broadcast with network coding. Optimal handover decision is usually based on the signal strength of the AP, however at the border contours of the different levels of the received signal to noise ratio where the decision can only be taken based on the path loss, we cannot guarantee that the station's service  will stay with the same quality. Therefore, introducing another objective is worth to think about. We introduce a framework that decides when and where is the optimal point(s) to do a network-coded handover. We need to optimize the handover decision to guarantee maximum received sum rates from AP1 the station is already accessing to AP2 the one the mobile station will handover to. Suppose that probing, authentication, and re-association times can be minimized, see \cite{macHandoff1},\cite{macHandoff2} where the authors of the first showed that probing contribute to the main delay in a handover process, while the later proposed a selective scanning and cashing mechanism to reduce the probing delay. This is in fact a forthcoming result to a decision based on the maximum sum rate objective. Therefore, the mobile station can be at anytime receiving from AP1 or AP2 before it becomes fully served by the AP it will be completely associated with and under its coverage far from the edge. This is a relevant assumption due to the fact that the station can experience ups and downs in the received $snr$ from different APs due to mobility and distance changes, while probing is done via broadcast messages which completes the setup introduced. The objective is formulated to maximize the rate cost functions in order to find the optimal switching time as follows:
\begin{equation}
\max_{t} \int_{t1}^{t} \! R_1(x(t),y(t)) \, \mathrm{d}t+ \int_{t+\tau}^{t2} \! R_2(x(t),y(t)) \, \mathrm{d}t   
\end{equation}
With,
\begin{equation}
R_1=\frac{N_{c~AP1}}{\mathbb{E}\left[T_{N_{c}~packets~AP1}\right]}
\end{equation}
and,
\begin{equation}
R_2=\frac{N_{c~AP2}}{\mathbb{E}\left[T_{N_{c}~packets~AP2}\right]}
\end{equation}
The point $x(t),y(t)$ corresponds to the coordinates of the station at a given time. And $t_1$ is the initial time at the beginning of the mobile station path, $\tau$ is the time to re-associate to AP2, and $t_2$ is the time of measurement at the end of the mobile station path.
%\Huge
%\setcounter{secnumdepth}{0}
\subsection{Station Associated to Both APs}
\normalsize
Worth to notice that we are interested to find optimal time $t^{\star}$ when the sum-rate can be maximized at which both stations can start to code across the packets via a broadcast transmission mode at the same time to let the handover process being performed with no distruption of the user experience, and the user will always find the required packet flows to decode, cashing can also be useful in implementing the setup introduced here to reduce the time of probing and full re-association. In particular, the network-coded handover introduced include the following steps: 
\begin{enumerate}
	\item When the signal strength received at a certain point is almost equivalent from both APs, the station starts sending probe messages to neighbor access points, and through cashing this process delay will be minimal, therefore, it will find AP2 as the first option in the neighboring list. 
\item The station will transmit to both APs via ACK messages the degrees of freedom it requires to continue the service without service disruption.
%\newline ~~{III-} 
\item AP1 will activate network-coded broadcast and AP2 will authenticate and associate the station before it fully disconnects from the previous AP1 and directly sends the network-coded broadcast data required. 
%\newline ~~{IV-} 
\item The station will continue to acknowledge its degrees of freedom to both APs, and by the mobility considerations, the station will be reassociated to AP2 while still receiving its coded-packets, when the association is completed to AP2. The station is not anymore connected to AP1. 
%\newline
\end{enumerate}
Therefore, we introduce a framework that would allow the station to be connected to two stations at a certain point receiving from both of them. The station will be announced at the optimal distance $d$ and so at the optimal time $t^{\star}$ to receive coded broadcast before it fully re-associate to the new AP, and so the network-coded handover will be reliably performed over broadcast while the station is moving; guaranteeing a near optimal packet flow without disruptions. In fact, the optimal time to switch to network-coding broadcast mode is an optimal set of points starting few meters back and few meters forth in the edges of both APs, the so called borders of the SNR contours of APs coverage, which are also path dependent. In other words, this point in time can be moved back if AP cooperation is implemented over backhaul link - with CSI and data sharing - similar to the framework in \cite{samahMCP}, while the exact optimal time when to switch to the new AP can be found numerically. Such framework is practically feasible in FiWi networks with optical fiber backhaul connecting the two APs to a central unit. Notice also that, one possible optimal point to start the transmission of coded packets from AP1 is at the distance corresponding to $d(t^{\star})-d(\tau)$ meters from the optimal distance to switch to AP2, $d(t^{\star})$, where by the time $t^{\star}$, the station will switch to AP2, receiving the required/lost DoF or innovative (new) packets. 

%\Huge
%\setcounter{secnumdepth}{0}
\subsubsection{Station Associated to One AP at a Time}
\normalsize
Here we consider that the path of the mobile station is known before hand, and so, given the location of the station $(x(t),y(t))$ at a given time, we can measure the distance from AP1 located at $(x_1,y_1)$ to the mobile station as follows:
\begin{equation}
d_1(t)=\sqrt{(x_1-x(t))^2+(y_1-y(t))^2}
\end{equation}
Similarly, we can measure the distance from AP2 located at $(x_2,y_2)$ to the mobile station as follows:
\begin{equation}
d_2(t)=\sqrt{(x_2-x(t))^2+(y_2-y(t))^2}
\end{equation}
Substituting the distances $d_1$ and $d_2$ of each AP from the mobile station into the objective that aims to maximize the sum rate received by the station, we can find the optimal set of distances where both APs can activate the network-coded mode of operation during which a soft handover process can take place seamlessly with least interruption over the service. It is not possible to find a closed-form expression for the optimal times to start coded packets transmission, or the optimal time $t^{\star}=d^{\star}/v$ to switch to the new AP.%t^{\star}+\tau$ to reassociate to the new AP, but we dont want to strongly emphasize it this way

Therefore, we can consider $R_1$ and $R_2$ using Shannon's capacity formula, this will provide more feasible way to find the optimal time to switch to the new AP, therefore, we can re-write the objective function as follows:
\begin{multline}
\max \int_{t_1}^{t} \! log(1+snr(x(t),y(t))) \, \mathrm{d}t \\
+ \int_{t+\tau}^{t_2} \! log(1+snr(x(t),y(t))) \, \mathrm{d}t   
\end{multline}
Deriving the integral of the cost function with respect to the distance and plugging the distance-time-velocity relation into the result, we can find the optimal time $t^\star$ to switch to the new AP, as the numerical solution of the equation above. However, we can account for more delays via AP cooperation and coding across packets.
%
%Deriving the integral of the cost function with respect to the distance and plugging the distance-time-velocity relation into the result, the objective becomes as follows:
%\renewcommand{\theequation}{9.41}
%\small
%\begin{multline}
%max \frac{vt_1}{2}ln(1+\frac{1}{t_1^{2}v^{2}})+  \frac{1}{2vt_1}ln(1+t_1^{2}v^{2}) -\frac{1}{2vt_1} \\
       %-\frac{vt}{2}ln(1+\frac{1}{t^{2}v^{2}})-  \frac{1}{2vt}ln(1+t^{2}v^{2}) +\frac{1}{2vt} \\
       %+\frac{v(t+\tau)}{2}ln(1+\frac{1}{(t+\tau)^{2}v^{2}})+  \frac{1}{2v(t+\tau)}ln(1+(t+\tau)^{2}v^{2}) -\frac{1}{2v(t+\tau)} \\
       %-\frac{vt_2}{2}ln(1+\frac{1}{t_2^{2}v^{2}})-  \frac{1}{2vt_2}ln(1+t_2^{2}v^{2}) +\frac{1}{2vt_2} \\
%\end{multline}
%\normalsize
%The derivative of the integral form with respect to the distance, and in terms of time-velocity, is as follows:
%\small
%\begin{multline}
        %\frac{v}{2}ln(1+\frac{1}{{t_1}^2{v}^2})- \frac{1}{2v{t_1}^2}ln(1+{t_1}^2{v}^2) +\frac{1}{2v{t_1}^2} \\
        %-\frac{v}{2}ln(1+\frac{1}{{t}^2{v}^2})+ \frac{1}{2v{t}^2}ln(1+{t}^2{v}^2) -\frac{1}{2v{t}^2} \\
        %+\frac{v}{2}ln(1+\frac{1}{{(t+\tau)}^2{v}^2})- \frac{1}{2v{(t+\tau)}^2}ln(1+{(t+\tau)}^2{v}^2) +\frac{1}{2v{(t+\tau)}^2} \\
        %-\frac{v}{2}ln(1+\frac{1}{{t_2}^2{v}^2})+ \frac{1}{2v{t_2}^2}ln(1+{t_2}^2{v}^2)-\frac{1}{2v{t_2}^2} \\
        %=0\\
%\end{multline}
%\normalsize
%Therefore, the optimal time to switch $t^\star$ is the numerical solution of the equation above. However, we can account for more delays via AP cooperation and coding across packets.
%\Huge
%\setcounter{secnumdepth}{0}
\section{ Simulation Results}
The expected time to deliver one or $N$ packets derived in the sequel of the previous sections can be adapted to different WiFi IEEE 802.11 systems. Table 2 provides a summary of the approved timing parameters in the specifications of the MAC-layer which decomposes the packets. The following simulations focus on the values of the IEEE 802.11g with legacy, using ${T}_{p}=0.00144~sec$. 

\begin{figure}
     \begin{center}
    \includegraphics[width=3.2in, height=2.4in]{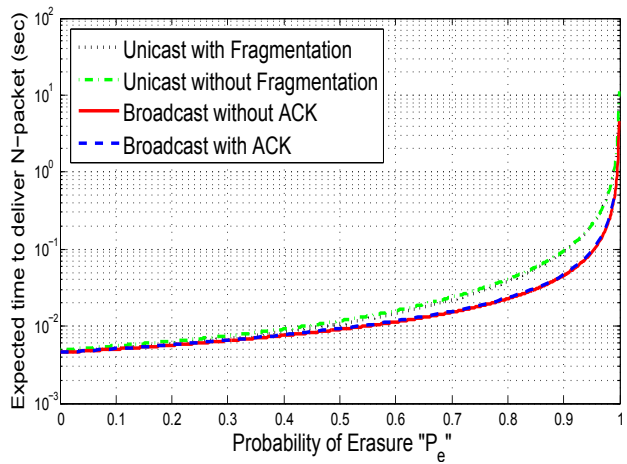}
    \caption{Time to deliver $N$ packets(sec) vs. probability of erasure.}
			\label{Fig:figure-7}
    \end{center}
    \label{figure-7}
\end{figure}

Figure~\ref{Fig:figure-7} illustrates the time to deliver $N$ packets by unicast, broadcast, and broadcast with ACK with respect to the probability of erasure. It is clear that the least delay incurred by using a broadcast mode under any number of coded or uncoded packets. 

\begin{figure}
     \begin{center}
    \includegraphics[width=3.2in, height=2.4in]{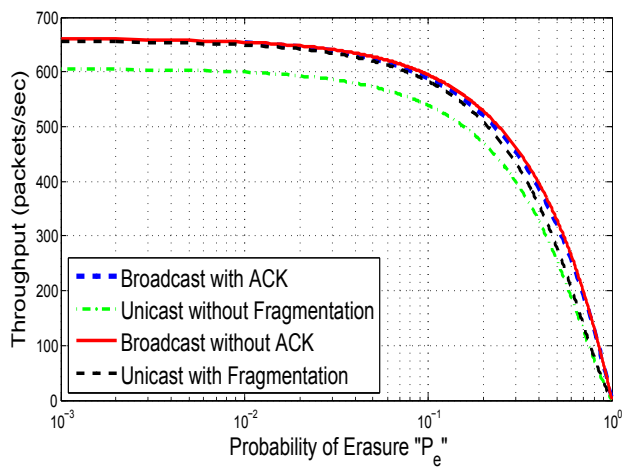}
    \caption{Throughput(packets/sec) vs. probability of erasure.}
			\label{Fig:figure-8}
    \end{center}
    \label{figure-8}
\end{figure}

Consequently, we can see in Figure~\ref{Fig:figure-8} that the Throughput(packets/sec) is higher for the broadcast transmission mode than for the unicast while broadcast with ACK is a reliable alternative to the one without ACK. The previous results are emphasized through the relation between the packet error rate and the bit error rate, which makes it possible to derive the time to deliver $N$ packets and the throughput with respect to the SNR and so with respect to the distance of the mobile station. 

\begin{figure}
     \begin{center}
    \includegraphics[width=3in, height=2.1in]{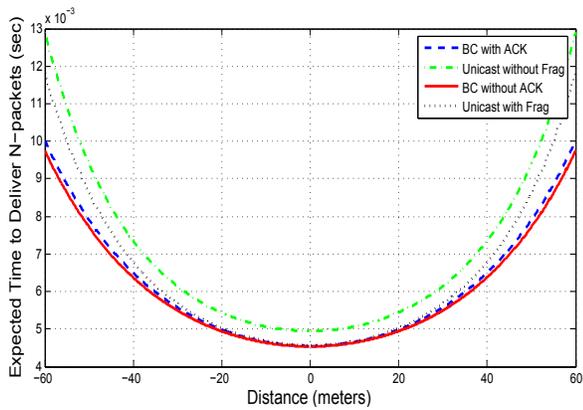}
    \caption{Time to deliver $N$ packets(sec) vs. distance(meters).}
			\label{Fig:figure-9}
    \end{center}
    \label{figure-9}
\end{figure}

\begin{figure}
   \begin{center}
    \includegraphics[width=3.2in, height=2.1in]{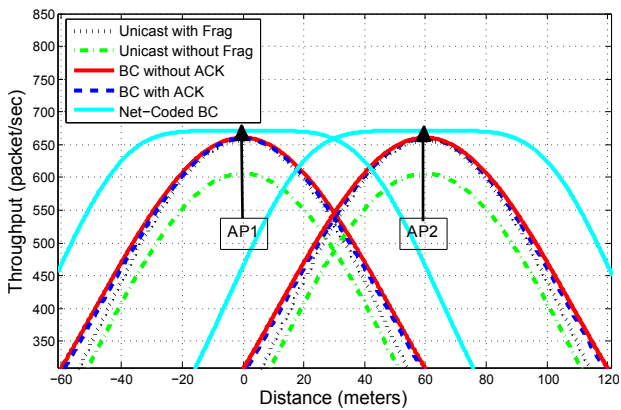}
    \caption{Throughput(packets/sec) vs. distance(meters).}
		\label{Fig:figure-10}
    \end{center}
    \label{figure-10}
\end{figure}

Figure~\ref{Fig:figure-9} illustrates the time to deliver the $N$ packets using different transmission modes with respect to the distance, and Figure~\ref{Fig:figure-10} illustrates the Throughput using different transmission modes with respect to the distance from both AP1 and AP2. For the sake of clarity, we suppose that the distance between AP1 and AP2 is 60 meters, i.e., AP1 is located at (0,0) and AP2 at (0,60). The negative distances correspond to the distances between the station and the access point at the other side of the coordinate. 

At a distance $d=1$ meters for an indoor scenario, the SNR maintained is 64.124 dB, at a distance of $d=32$ meters indoor or outdoor ($1000>d>100$) the maintained SNR is 42.11 dB. Therefore, we can chose $c=100000$ which is an acceptable figure when a fading scenario is assumed. 

We can see that the broadcast transmission always outperform the unicast one. However, we can also see the gain of introducing coding across the packets where the transmission of network-coded packets introduces a gain at earlier distances and with proper tuning to the system parameters, the throughput via network coding can approach the theoretical capacity of broadcast channels. Proper tuning means that we can limit the number of retransmissions of a lost packet, this is of practical relevance since the packets have a timeout that afterwards its worthless for the receiver. 

To illustrate the optimal decision set of a network-coded handover corresponding to the optimal time or distance at which the mobile station can receive coded packets from both current and prospective AP, and during the coded handover on the edge process. Suppose that AP1 is located at $(0,0)$ and AP2 at $(25,0)$, and the path the mobile station is moving along, is deterministic, and given by a linear movement with $y(t)=x(t)$, the maximum of the sum rates can be received by the mobile at a specific (or several) points in time. 

However, due to the fact that the optimal time is not only distance but also velocity dependent $t^{\star}=d^^*/v$; if the mobile is moving with constant speed equals $v=2$ meters/sec and the delays of switching to AP2 is $\tau=10sec$. $t^\star\in[t_1:t_2]$ corresponds to the optimal time to switch $t^{\star}=11.5~sec$ which corresponds to time when the station is associated to AP2.

This means that the optimal distance at which it is optimal to handoff to the new AP is equal to $d_2(t^{\star})=t^{\star}v=23$ meters; which is path and velocity dependent. This also defines a possible point in time where AP1 can start transmission of its coded packets, this may be possible at $d(t^{\star})-d(\tau)=3$ meters before the new association takes place which corresponds to $d_1(t)=23-3=20$ meters.

If the path of travel of the mobile station is partially or totally unknown, location estimation techniques and some mobility models can be used instead, depending on the network topology under study.
\vspace{-1cm}
\vspace{1cm}
\vspace{1cm}
\begin{table}[ht]
\caption{Timing parameters of the IEEE 802.11 MAC layer}   % title of Table
\centering % used for centering table
\begin{tabular}{| c | c | c | c | c |}  % centered columns (5  columns)
 \hline\hline
  Parameter & 802.11b & 802.11a & 802.11g & 802.11g+legacy\\
 %           &         &         &         &\\
  \hline
  $T_{slot}$ & $20\mu{s}$  & $9\mu{s}$ & $9\mu{s}$ & $20\mu{s}$\\
  \hline
  $SIFS$ & $10\mu{s}$  & $16\mu{s}$ & $10\mu{s}$ & $10\mu{s}$\\
   \hline
  $DIFS$ & $50\mu{s}$  & $34\mu{s}$ & $28\mu{s}$ & $50\mu{s}$\\
  \hline
  $ACK$ & $14{Bytes}$  & $14{Bytes}$ & $14{Bytes}$ & $14{Bytes}$\\
  \hline
  $CW_{min}$ & $31T_{slot}$  & $15T_{slot}$ & $15T_{slot}$ & $15T_{slot}$\\
  \hline
  $CW_{max}$ & $1023T_{slot}$  & $1023T_{slot}$ & $1023T_{slot}$ & $1023T_{slot}$\\
  \hline
%\end{tabular*}
\end{tabular}
\label{table1}
\end{table}
\section{ Conclusion}
In this paper, we propose a new framework that considers network coding on the edge as a key enabling technique to apply in 5G wireless networks with stringent delay requirements and small cell densification. This potential of this technique resorts to its feasibility with different operational modes with or without AP cooperation and for different mobile user equipment if it supports dual or single mode. To address this proposal, we propose a set of novel models of the DCF of the WiFi IEEE 802.11 with fixed time contention window for unicast transmission, with and without fragmentation, and for the uncoded and coded broadcast transmission, with and without ACK. We analyze the delay over the unicast and broadcast transmission for a network topology that includes one AP and one station. We provide closed-form expressions for the expected time to deliver the $N$ packets for the DCF mechanism, with  unicast, the general broadcast, broadcast with ACK for the uncoded and coded transmission. We have shown that coding across packets in an acknowledged broadcast scenario encounters less delays, higher reliability, and higher throughout than for the uncoded broadcast or unicast cases. We propose a new protocol that utilizes network coding to broadcast coded packets to the station performing handover. This new proposed network-coded on the edge handover framework will immensely serve if implemented in the current standardized IEEE 802.11 systems. We build upon constraints that take into consideration the distance of the station and the degrees of freedom it owns to be able to decode the received packets before it switches the connection to the new AP. This has been demonstrated by a novel mathematical formulation of a network-coded handover on the edge that decides the optimal time when to switch to the new access point in order to maximize the sum rates received by both stations. In particular, such optimal time to do a handover between two APs accounts for the delay penalty due to probing, association and authentication to the new AP. In addition, we propose a framework were the mobile can be associated to two APs simultaneously and so coding across packets can be cooperated between the two APs to keep the flow of native (uncoded) packets and account for lost ones in cooperative fashion mimiking the a coded repition framework for higher reliability. Therefore, we provide a framework under which the QoS over delay sensitive streaming applications can be radically improved, providing a seamless handover. This in conjunction with the techniques proposed underly a proposal on adding a network coding layer on top of the classical MAC layer which can improve the current standardized IEEE 802.11 and boost its performance from a delay and throughput perspective. This will be of particular relevance in delay sensitive applications if implemented in existing technologies. Worth to note that it is intuitive that the addition of network coding can boost the data rates, and minimize the delay. However, NC adds a decoding complexity if we consider a large set of users with high mobility. In such scenarios, a conservative approach can be thought of, in order to provide an optimal decision of when to switch on/off the network coding. This can be applied to mobile nodes based on their SNR levels, whether they are performing a handover or just losing the quality due to mobility or obstructions. Nevertheless, its instrumental to think of a NC layer as an enhancement layer that can be, adapted to different set of parameters, like the physical parameters of the communication channel, or the number of mobile stations associated to the APs. In addition, a NC layer can be adapted to solve a set of problems, like performing a seamless handover, coping with unbalanced data rate users demands, or solving the issues encountered by mobile stations experiencing poor coverage. In principle, a NC layer is worth to be implemented and verified in current technologies, with a centralized control mechanism via the APs on its adaptation, or with distributed control mechanisms via associated stations, i.e., like having on demand service. Future work will consider the multi-station modeling problem, the speed of the mobile station, and with mobility patterns or unknown paths of the mobile station. 
\bibliographystyle{IEEEtran}
% argument is your BibTeX string definitions and bibliography database(s)
\bibliography{IEEEabrv,mybibfile}
% <OR> manually copy in the resultant .bbl file
% set second argument of \begin to the number of references
% (used to reserve space for the reference number labels box)
%\begin{thebibliography}{1}
%\bibitem{IEEEhowto:kopka}
%H.~Kopka and P.~W. Daly, \emph{A Guide to \LaTeX}, 3rd~ed.\hskip 1em plus
%  0.5em minus 0.4em\relax Harlow, England: Addison-Wesley, 1999.

%\end{thebibliography}

% that's all folks
\end{document}